\documentclass[prd ]{revtex4}%
\usepackage{caption}
\usepackage{amsmath}
\usepackage{graphicx}
\usepackage{epsfig}
\usepackage{dcolumn}
\usepackage{bm}
\usepackage{slashed}
\usepackage{amsfonts}
\usepackage{amssymb}%
\providecommand{\U}[1]{\protect\rule{.1in}{.1in}}
\begin{document}
\title{ About the effects of diffusion in electric tomography }
\author{Alejandro Cabo$^{*}$, Jorge J. Riera$^{**}$ and Pedro A. Vald\'es
Hern\'andez$^{***}$ \bigskip}

\address{$^*$ Department of  Theoretical Physics, Instituto de Cibern\'etica, Matem\'atica y F\'isica ICIMAF, La Habana, Cuba. \bigskip\\
$^{**}$ Department of Biomedical Engineering, Florida International University, Miami, Florida. \bigskip \\
$^{**}$ Departamento de Neuroimagenes,\\ Centro de Neurociencias de Cuba, La Habana, Cuba }

\begin{abstract}
\noindent The Maxwell equations for an homogeneous medium in which electric
currents are composed of Ohmic, diffusive and impressed currents are written
in the static approximation in which displacements currents are neglected.
Closed exact expressions are given for the solutions of the charge density
$\rho,$ electric field potential and the magnetic field intensity as functions
of the impressed currents and densities ($\overrightarrow{J}_{imp}$,
$\rho_{imp}$). Also, approximate formulae for these magnitudes are derived.
The solutions correctly reproduce the conclusions of a previous work in which
the relevance of Yukawa like potentials determined by diffusion effects were
identified. Within the approximate solutions, the effective potential
generated by the impressed sources is given by the superposition of Yukawa
potentials created by the impressed charges concentrated in a given volume
element. This representation makes evident that at small spatial regions, the
experimental detectors are able to measure spherically symmetrical signals,
indicating the presence of electric monopoles. The discussion also allows to
argue that the magnetic field tomographic signal becomes completely
independent of the presence of diffusion in the medium. The analysis also
permits to determine the range of space and temporal frequencies in which the
diffusion effects might become important in justifying the detection of
monopole signals.

\end{abstract}
\maketitle

\section{Introduction}

To clarify the physical role of the diffusion effects in current EEG within
small spatial dimensions is a topic of relevance in neurophysiology
\cite{riera1,riera2,riera3, otros1,otros2,otros3,otros4}. The work
\cite{riera1} represented a breakthrough in the research activity in this
theme. The results indicated the existence of electric monopole components in
the current charge densities measured from the tomography of very small
samples within mouse brains. After that, the origin of these effects had been
intensely debated \cite{riera1,riera2,riera3,otros1,otros2,otros3,otros4}. In
reference \cite{riera3} it was considered the numerical solutions of a set of
equations for the charge density in an homogeneous model of the brain
including Ohmic as well as diffusive currents in addition to the impressed
sources. The results indicated the possibility for detecting spherically
symmetric solutions when the electrodes are of the order of distance
$\frac{\epsilon d}{\sigma a}$ where the $\epsilon$ is the dielectric constant
of the tissue, $d$ is the diffusion constant, $\sigma$ is the tissue
conductivity and $a$ is typical size of the spatial region in which the
impressed sources are concentrated.

In the present work we present and solve the equations for the model
investigated in \cite{riera3}. Firstly, we derive the set of Maxwell equations
by fixing the approximation in which the displacement current contribution to
the electric field is disregarded. The linear equations for the charge
density, the electric potential and the magnetic field are solved exactly.
Afterwards, it is performed an approximation which strongly simplifies the
solutions for the frequencies usually measured in electroencephalography when
the characteristic frequency $\sqrt{\frac{\sigma}{\epsilon}}$ and the
frequency bandwidth of the measured signals $\Delta w$ satisfies $\Delta
w\sqrt{\frac{\sigma}{\epsilon}}\ll1$. The approximation corresponds to
disregard the time in which the final static Yukawa solution is established
after suddenly adding a free point charge in the medium. After this, the
solution for the potential generated by completely general form of impressed
currents is expressed as the superposition of Yukawa potentials centered in
each element of volume and weighted by the impressed charge density at this
point. This expression directly implies that when measures are done at spatial
distances of the order of $\frac{\epsilon d}{\sigma}$ , it is possible to
detect spherically symmetric potentials indicating the presence of a net
charge (electric monopoles) in the interior regions. Further, the solution for
the magnetic field is also derived. The result is simpler: the diffusion in no
way affects the magnetic encephalography signals. We also investigate the
range of spatial regions in which monopole signals could be most easily
detected. It follows that for dipole sizes in the range 0-10 $\mu m$ , the
dipole potential has a ratio with respect to the monopole potential, being
smaller than the number 6 which is estimated from the results in
\cite{riera1}, for some possible values of the brain tissue parameters.

In section 2, we derive the set of Maxwell equations in the static
approximation. Section 3 is devoted to expose the solution for the electron
charge density. First, the exact solution for the density is written in terms
of a Green function for the obtained linear equation, acting as a kernel over
the charge density. In addition, the approximate expression for the solution
is also derived here. The section 4 consider the determination of the total
electrostatic potential also as a linear expression of the impressed sources.
In the next section 5, the solution for the magnetic field is presented.
Section 6 discuss the dimensions of the spatial regions in which monopole
signals could be observed in dependence of the parameters of the brain tissue.
Finally, the results are reviewed in the summary in last section.

\section{Maxwell equations for an homogeneous brain model}

Let us present in this section the basic equations to be employed in
describing the homogeneous model of the brain tissue including Ohmic,
diffusion and impressed currents. Since the static approximation will be used,
the Gauss Law equations becomes
\begin{equation}
-\nabla^{2}\phi=\frac{1}{\epsilon}\rho+\frac{1}{\epsilon}\rho_{imp},
\end{equation}
where $\rho_{imp}$ is the impressed charge densities assumed to be created by
the neurons through their active covering membranes, $\rho$ is the total
charge density in the brain medium being external to the cells $\phi$ is the
electric potential which negative gradient gives the electric field intensity.
Further, the total electric current will be written in the form
\begin{equation}
\overrightarrow{J}_{T}=-(\sigma\overrightarrow{\nabla}\phi+d\text{
}\overrightarrow{\nabla}\rho)+\overrightarrow{J}_{imp},
\end{equation}
in which the first term is the Ohmic current, the second is the current
associated to the diffusion (given as the negative of the gradient of the
density $\rho_{imp}$) and the last term is the impressed current created at
the neuron's surfaces. Let us assume now that the impressed current and
sources satisfy the local charge conservation condition
\begin{equation}
0=\frac{\partial}{\partial t}\rho_{imp}+\overrightarrow{\nabla}\cdot
\overrightarrow{J}_{imp}.
\end{equation}
Then, the conservation condition for the total current $\overrightarrow{J}%
_{T}$ and the total density $\rho+\rho_{imp}$ leads to
\begin{align}
\overrightarrow{\nabla}\cdot\overrightarrow{J}_{T}+\frac{\partial}{\partial
t}(\rho+\rho_{imp})  &  =0,\\
\frac{\partial}{\partial t}\rho+\frac{\sigma}{\epsilon}\rho-d\text{ }%
\nabla^{2}\rho+\frac{\sigma}{\epsilon}\rho_{imp}  &  =0,
\end{align}
which is a closed equation for the determination of the volume charge density
$\rho$ as a functional of the impressed charge density $\rho_{imp}.$ The
electric and magnetic fields are defined as usual by
\begin{align}
\overrightarrow{E}  &  =-\overrightarrow{\nabla}\phi,\text{ \ }\\
\text{\ }\overrightarrow{B}  &  =\overrightarrow{\nabla}\times
\overrightarrow{A}.
\end{align}

The Ampere's Law can be written as
\begin{align}
\overrightarrow{\nabla}\times\overrightarrow{B}  &  =\overrightarrow{J}%
_{T}+\epsilon\frac{\partial}{\partial t}\overrightarrow{E}\nonumber\\
&  =\overrightarrow{J}_{T}-\epsilon\frac{\partial}{\partial t}%
\overrightarrow{\nabla}\phi,
\end{align}
in which the displacement current $\epsilon\frac{\partial}{\partial
t}\overrightarrow{E}$ should not be neglected in order to be able to satisfy
the local current conservation condition%
\begin{align}
\overrightarrow{\nabla}\cdot\overrightarrow{J}_{T}+\epsilon\frac{\partial
}{\partial t}\overrightarrow{\nabla}\cdot\overrightarrow{E}  &  =0,\nonumber\\
\overrightarrow{\nabla}\cdot\overrightarrow{J}_{T}+\frac{\partial}{\partial
t}\rho+\frac{\partial}{\partial t}\rho_{imp}  &  =0.
\end{align}

The set of equations for the determination of the density $\rho,$ the electric
potential $\phi$ and the magnetic field in terms of the impressed sources
generating them, can be summarized as follows%
\begin{align}
-\nabla^{2}\phi &  =\frac{1}{\epsilon}\rho+\frac{1}{\epsilon}\rho_{imp},\\
0  &  =\frac{\partial}{\partial t}\rho+\frac{\sigma}{\epsilon}\rho-d\text{
}\nabla^{2}\rho+\frac{\sigma}{\epsilon}\rho_{imp} ,\label{density}\\
\overrightarrow{\nabla}\times\overrightarrow{B}  &  =-(\sigma
\overrightarrow{\nabla}\phi+d\text{ }\overrightarrow{\nabla}\rho
)+\epsilon\frac{\partial}{\partial t}\overrightarrow{E}+\overrightarrow{J}%
_{imp},\\
\overrightarrow{E}  &  =-\overrightarrow{\nabla}\phi,\text{ \ \ }%
\overrightarrow{B}=\overrightarrow{\nabla}\times\overrightarrow{A},\\
0  &  =\frac{\partial}{\partial t}\rho_{imp}+\overrightarrow{\nabla}%
\cdot\overrightarrow{J}_{imp}.
\end{align}

\section{Solution for the charge density}

\ The equation for the density (\ref{density}),
\[
\frac{\partial}{\partial t}\rho+\frac{\sigma}{\epsilon}\rho-d\text{ }%
\nabla^{2}\rho=-\frac{\sigma}{\epsilon}\rho_{imp}%
\]
can be solved by acting on it on the left with the inverse kernel $G$ of the
differential operator $\frac{\partial}{\partial t}+\frac{\sigma}{\epsilon}-d$
$\nabla^{2}$, leading to
\begin{equation}
\rho(\overrightarrow{x},t)=-\frac{\sigma}{\epsilon}\int d\overrightarrow{x}%
^{\prime}dt^{\prime}\text{ }G(\overrightarrow{x},t\text{ };\overrightarrow{x}%
^{\prime},t^{\prime})\rho_{imp}(\overrightarrow{x}^{\prime},t^{\prime
})\text{,} \label{rho}%
\end{equation}
in which the mentioned kernel can be written in the Fourier representation as
\begin{equation}
G(\overrightarrow{x},t\text{ };\overrightarrow{x}^{\prime},t^{\prime}%
)=\int\frac{d\overrightarrow{k}dw}{(2\pi)^{4}}\text{ }\frac{\exp(i\text{
}(\overrightarrow{x}-\overrightarrow{x}^{\prime})\cdot\overrightarrow{k}%
-i\text{ }(t-t^{\prime})w)}{-i\text{ }w+\frac{\sigma}{\epsilon}+d\text{
}\overrightarrow{k}^{2}}.
\end{equation}

But, the following integral identity can be employed
\[
\int i\text{ }dw\frac{\exp(-i\text{ }\tau\text{ }w)}{w+i(\frac{\sigma
}{\epsilon}+d\text{ }\overrightarrow{k}^{2})}=2\pi\text{ }\theta(\tau)\text{
}exp(-(\frac{\sigma}{\epsilon}+d\text{ }\overrightarrow{k}^{2})\tau),
\]
where $\theta(\tau)$ is the Heaviside unit step function\cite{bateman}. This
allows to rewrite for $G$ the expression
\begin{equation}
G(\overrightarrow{x},t\text{ };\overrightarrow{x}^{\prime},t^{\prime})=\text{
}\theta(t-t^{\prime})\exp(-\frac{\sigma}{\epsilon}(t-t^{\prime}))\int%
\frac{d\overrightarrow{k}}{(2\pi)^{3}}\text{ }\exp(i\text{ }%
(\overrightarrow{x}-\overrightarrow{x}^{\prime})\cdot\overrightarrow{k}%
-d\text{ }\overrightarrow{k}^{2}(t-t^{\prime})).
\end{equation}

\ In the case of a very rapid decaying with the time difference $t-t^{\prime}$
of $G$ allows to determine a solution representing a very good approximation
for the exact charge density $\rho$. For this purpose let us express the
integral over $t^{\prime}$ in (\ref{rho}) (after substituting the Fourier
expansion for $G$) by an equivalent integral over the difference
$\lambda=(\frac{\sigma}{\epsilon}+d$ $\overrightarrow{k}^{2})(t-t^{\prime})$
in the form%
\begin{align}
\rho(\overrightarrow{x},t)  &  =-\frac{\sigma}{\epsilon}\int
d\overrightarrow{x}^{\prime}dt^{\prime}\text{ }\theta(t-t^{\prime})\exp
(-\frac{\sigma}{\epsilon}(t-t^{\prime}))\times\\
&  \int\frac{d\overrightarrow{k}}{(2\pi)^{3}}\text{ }\exp(i\text{
}(\overrightarrow{x}-\overrightarrow{x}^{\prime})\cdot\overrightarrow{k}%
-d\text{ }\overrightarrow{k}^{2}(t-t^{\prime}))\rho_{imp}(\overrightarrow{x}%
^{\prime},t^{\prime})\\
&  =-\frac{\sigma}{\epsilon}\int d\overrightarrow{x}^{\prime}\int%
\frac{d\overrightarrow{k}}{(2\pi)^{3}}\exp(i\text{ }(\overrightarrow{x}%
-\overrightarrow{x}^{\prime})\cdot\overrightarrow{k})\times\\
&  \int_{0}^{\infty}\frac{d\lambda}{\frac{\sigma}{\epsilon}+d\text{
}\overrightarrow{k}^{2}}\exp(-\lambda)\rho_{imp}(\overrightarrow{x}^{\prime
},t-\frac{\lambda}{\frac{\sigma}{\epsilon}+d\text{ }\overrightarrow{k}^{2}}).
\end{align}

\ But, it can be noticed when the screening time $\frac{\epsilon}{\sigma}$ is
very much larger than the unit, the integrand over the dimensionless quantity
$\lambda$ should be concentrated near the values of $\lambda$ being of the
order of the unit. This is because in this case $\rho_{imp}(\overrightarrow{x}%
^{\prime},t-\frac{\lambda}{\frac{\sigma}{\epsilon}+d\text{ }\overrightarrow{k}%
^{2}})\rightarrow\rho_{imp}(\overrightarrow{x}^{\prime},t)$ becomes
independent of $\lambda.$ In this approximation, it follows
\begin{equation}
\rho(\overrightarrow{x},t)=-\int\frac{d\overrightarrow{k}}{(2\pi)^{3}}%
\frac{\exp(i\text{ }\overrightarrow{x}\cdot\overrightarrow{k})}{1+\frac
{d\text{ }\epsilon}{\sigma}\overrightarrow{k}^{2}}\rho_{imp}%
(\overrightarrow{k},t).
\end{equation}
which defines an approximate solution for the volumetric density $\rho$ as a
function of the impressed density. \

\bigskip

\section{Solution for the electric potential}

The solution determining the electric tomographic signals is the one
corresponding to the electric potential. It can be derived by using the
already obtained solution for the volume charge density%
\begin{equation}
\rho(\overrightarrow{x},t)=-\frac{\sigma}{\epsilon}\int d\overrightarrow{x}%
^{\prime}dt^{\prime}\text{ }G(\overrightarrow{x},t\text{ };\overrightarrow{x}%
^{\prime},t^{\prime})\rho_{imp}(\overrightarrow{x}^{\prime},t^{\prime})\text{
}%
\end{equation}

Then, the Gauss Law in differential form%
\begin{equation}
-\nabla^{2}\phi=\frac{1}{\epsilon}\rho+\frac{1}{\epsilon}\rho_{imp},
\end{equation}
and the inverse kernel of the Laplacian allows to write the potential as
follows%
\begin{equation}
\phi(\overrightarrow{x},t)=\int d\overrightarrow{x}^{\prime}\text{ }\frac
{1}{4\pi|\overrightarrow{x}-\overrightarrow{x}^{\prime}|}(\frac{1}{\epsilon
}\rho(\overrightarrow{x}^{\prime},t)+\frac{1}{\epsilon}\rho_{imp}%
(\overrightarrow{x}^{\prime},t)).
\end{equation}
By substituting the solution for the density, the exact expression for the
electric potential as a functional of the impressed sources becomes%
\begin{equation}
\phi(\overrightarrow{x},t)=\int d\overrightarrow{x}^{\prime}\text{ }\frac
{1}{4\pi\epsilon|\overrightarrow{x}-\overrightarrow{x}^{\prime}|}%
(-\frac{\sigma}{\epsilon}\int d\overrightarrow{x}^{\prime\prime}dt^{\prime
}\text{ }G(\overrightarrow{x}^{\prime},t\text{ };\overrightarrow{x}%
^{\prime\prime},t^{\prime})\rho_{imp}(\overrightarrow{x}^{\prime\prime
},t^{\prime})+\rho_{imp}(\overrightarrow{x}^{\prime},t)).
\end{equation}

\subsection{The approximate \ expression for the total potential}

Let us perform again the previous approximation in which the screening times
are assumed to be very much rapid, but in an alternative way. Remembering
that
\[
-\nabla_{\overrightarrow{x}}^{2}\frac{1}{4\pi|\overrightarrow{x}%
-\overrightarrow{x}^{\prime}|}=\delta^{(3)}(\overrightarrow{x}%
-\overrightarrow{x}^{\prime}),
\]
and acting with the operator $-\nabla_{\overrightarrow{x}}^{2}$ over
$\phi(\overrightarrow{x},t)$ permits to rewrite the known expression
\[
-\epsilon\nabla_{\overrightarrow{x}}^{2}\phi(\overrightarrow{x},t)=\rho
(\overrightarrow{x},t)+\rho_{imp}(\overrightarrow{x},t),
\]
which after Fourier transformed gives
\begin{align}
\epsilon\overrightarrow{k}^{2}\phi(\overrightarrow{k},w) &  =\rho
(\overrightarrow{k},w)+\rho_{imp}(\overrightarrow{k},w)\nonumber\\
&  =(1-\frac{1}{-i\text{ }\frac{\text{ }\epsilon}{\sigma}w+1+\frac{d\text{
}\epsilon}{\sigma}\overrightarrow{k}^{2}})\rho_{imp}(\overrightarrow{k},w).
\end{align}

But, in biological tissues the term $\frac{\sigma}{\epsilon}$ is a frequency
parameter much might be higher than the measured frequencies $w$ in usual
tomographic experiments. Thus, when the ratio $\frac{\text{ }\epsilon}{\sigma
}w$ might be very much smaller than the unit, ignoring  this term in the
denominator of the above expression, the Fourier transform of total value of
the electrostatic potential generated by the impressed densities can be
written as
\begin{align*}
\phi(\overrightarrow{k},w) &  =\frac{1}{\epsilon\overrightarrow{k}^{2}}%
(\rho(\overrightarrow{k},w)+\rho_{imp}(\overrightarrow{k},w))\\
&  =\frac{1}{\epsilon\overrightarrow{k}^{2}}(1-\frac{1}{1+\frac{d\text{
}\epsilon}{\sigma}\overrightarrow{k}^{2}})\rho_{imp}(\overrightarrow{k},w)\\
&  =\frac{\frac{d\text{ }}{\sigma}}{1+\frac{d\text{ }\epsilon}{\sigma
}\overrightarrow{k}^{2}}\rho_{imp}(\overrightarrow{k},w).
\end{align*}

This relation can be inversely Fourier transformed to the spatial
representation to write
\begin{align*}
\phi(\overrightarrow{x},t)  &  =\int d\overrightarrow{k}\text{ }dw\text{ }%
\phi(\overrightarrow{k},w)\exp(i\text{ }\overrightarrow{k}\cdot
\overrightarrow{x}-i\text{ }w\text{ }t)\\
&  =\int d\overrightarrow{k}\text{ }dw\text{ }\frac{\frac{d\text{ }}{\sigma}%
}{1+\frac{d\text{ }\epsilon}{\sigma}\overrightarrow{k}^{2}}\rho_{imp}%
(\overrightarrow{k},w)\exp(i\text{ }\overrightarrow{k}\cdot\overrightarrow{x}%
-i\text{ }w\text{ }t)\\
&  =\int d\overrightarrow{x}^{\prime}\text{ }\frac{1}{\epsilon
|\overrightarrow{x}-\overrightarrow{x}^{\prime}|}\exp(-\sqrt{\frac{\sigma
}{\epsilon\text{ }d}}\text{ }|\overrightarrow{x}-\overrightarrow{x}^{\prime
}|)\text{ }\rho_{imp}(\overrightarrow{x}^{\prime},t).
\end{align*}

This approximate solution gives the measured electrostatic potential as a
superposition of Yukawa potentials generated for the impressed charges
contained in each volume element. This conclusion reproduces in a generalized
way the results previously derived in reference \cite{riera3}. The expression
also tells that, under the assumed rapid screening approximation, the neural
impulses detected in the EEG experiments have dependences which are coincident
withe ones associated to the impressed sources. That is, there are no time
delay effects if the rapid screening occurs.

However, as noted above in biological tissues, at difference with usual
metals, the \ characteristic screening frequencies may lie in the interval
$0.00384148 $ to $0.119024$ seconds, which are by now mean higher than the
usual measured signals. This difference with usual metals occurs \ mainly
because the high values of the relative dielectric constant $\epsilon$ ranging
from $10^{5}$ to $10^{7}.$ Such large values are determined by macroscopic
properties associated to capacitive effect determined by the cell structure.
For example, for a usual medium like water, its relative dielectric function
determines a characteristic  frequency $\sqrt{\frac{\sigma}{\epsilon}}$
$=0.000119024$ $s$ which  for $w$ of order $100$ $Hz$ satisfies $w\sqrt
{\frac{\sigma}{\epsilon}}\ll1.$ \ Therefore, the approximated solution can be
employed only when brain tissue satisfies $w\sqrt{\frac{\sigma}{\epsilon}}%
\ll1.$ Thus, to employ the above approximation it should be verified that the
screening frequencies $\frac{\sigma}{\epsilon\text{ }}$ are very much larger
than the width of the frequency spectrum of the measured signals, that is
$w\sqrt{\frac{\sigma}{\epsilon}}\ll1.$

\section{ Solution for the magnetic field}

\ The solution for magnetic field intensity can be determined from the
Ampere's Law by transforming it as follows%
\begin{align}
\overrightarrow{\nabla}\times\overrightarrow{B}  &  =-(\sigma
\overrightarrow{\nabla}\phi+d\text{ }\overrightarrow{\nabla}\rho
)+e\frac{\partial}{\partial t}\overrightarrow{E}+\overrightarrow{J}%
_{imp}\nonumber\\
&  =\overrightarrow{\nabla}\times\overrightarrow{\nabla}\times
\overrightarrow{A}=-\nabla^{2}\overrightarrow{A}+\overrightarrow{\nabla
}(\overrightarrow{\nabla}\cdot\overrightarrow{A}),
\end{align}
which by after employing the Coulomb gauge $\overrightarrow{\nabla}%
\cdot\overrightarrow{A}=0,$ can be written as follows%
\begin{equation}
-\nabla^{2}\overrightarrow{A}=-\overrightarrow{\nabla}(\sigma\text{ }%
\phi+d\text{ }\rho+e\frac{\partial}{\partial t}\phi)+\overrightarrow{J}_{imp}.
\end{equation}
In the above expression, an important issue to notice is that the
non-impressed components of the current have gradient structures. After using
the inverse kernel for the Laplacian, the expression for the vector potential
can be written as
\begin{equation}
\overrightarrow{A}(\overrightarrow{x},t)=\int d\overrightarrow{x}^{\prime
}\text{ }\frac{1}{4\pi|\overrightarrow{x}-\overrightarrow{x}^{\prime}%
|}(-\nabla^{\prime}(\sigma\text{ }\phi(\overrightarrow{x}^{\prime},t)+d\text{
}\rho(\overrightarrow{x}^{\prime},t)+e\frac{\partial}{\partial t}%
\phi(\overrightarrow{x}^{\prime},t))+\overrightarrow{J}_{imp}%
(\overrightarrow{x}^{\prime},t)).
\end{equation}
Therefore, after taking into account the gradient nature of the non-impressed
components of the currents, it can be written the following solution for the
magnetic field intensity
\begin{align}
\overrightarrow{B}(\overrightarrow{x},t)  &  =\overrightarrow{\text{ }\nabla
}\times\int d\overrightarrow{x}^{\prime}\text{ }\frac{1}{4\pi
|\overrightarrow{x}-\overrightarrow{x}^{\prime}|}\overrightarrow{J}%
_{imp}(\overrightarrow{x}^{\prime},t)\nonumber\\
&  =-\int d\overrightarrow{x}^{\prime}\text{ }\frac{1}{4\pi|\overrightarrow{x}%
-\overrightarrow{x}^{\prime}|}\overrightarrow{\text{ }\nabla^{\prime}}%
\times\overrightarrow{J}_{imp}(\overrightarrow{x}^{\prime},t).
\end{align}
A central property to notice in this expression is that it is completely
independent of the presence of diffusion in the system. Thus diffusion does
not affect the magnetic tomographic images.

\section{Discussion about monopole detection within the model}

Let us further discuss the possibilities for the model of predicting the
observation of electric monopoles. This point was started to be analyzed in
reference \cite{riera3}. The discussion presented there is supported by the
results of this work. The signal for the 1 Coulomb value charge screened by
the Yukawa potential, but now written in International System of Units (ISU),
has the form
\begin{equation}
\phi_{m}(\overrightarrow{x}-\overrightarrow{x}^{\prime})=\text{ }\frac{1}%
{4\pi\epsilon\text{ }\epsilon_{0}|\overrightarrow{x}-\overrightarrow{x}%
^{\prime}|}.\exp(-\sqrt{\frac{\sigma^{ISU}}{\epsilon\epsilon_{0}\text{ }D}%
}\text{ }|\overrightarrow{x}-\overrightarrow{x}^{\prime}|),
\end{equation}
where $\ \epsilon_{0}=8.8542\times10^{-12}$ F/m is the vacuum permittivity
\ and $\epsilon$ is the relative dielectric constant and $\sigma^{ISU},D$ are
the electric conductivity and the diffusion constant respectively, but taken
in the ISU units. This solution can be interpreted as the stationary response
of the medium to the instantaneous implantation of a 1 C \ point charge at the
coordinates $\overrightarrow{x}^{\prime}$ . Let us call this solution as the
$monopole$ $potential$.

\begin{figure}[h]
\begin{center}
\includegraphics[width=8cm]{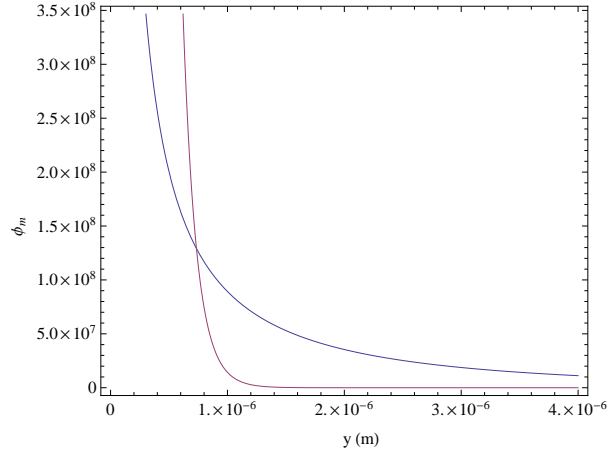}
\end{center}
\caption{ The spatial dependence of the stationary Yukawa potential generated
by 1 Coulomb point charge instantaneously inserted in the medium. Two plot are
shown: a) the top curve at large distances, is potential produced in a tissue
showing a largest values of the dielectric and diffusion constants and the
lower conductivity value, in the ranges defined in (\ref{1}),(\ref{2}%
),(\ref{3}). This determines the largest value for the Yukawa length. b) The
lower curve at large distances, related with the opposite situation in which
the conductivity and diffusion constant are the lowest values and the
conductivity the largest one. }%
\label{fig1}%
\end{figure}

Let us now compare the magnitude of the $monopole$ $potential$ with the one
associated to a $dipole$ $potential$, defined (for $|\overrightarrow{d}%
|\ll|\overrightarrow{x}-\overrightarrow{x}^{\prime}|$) as%
\begin{align}
\phi_{d}(\overrightarrow{x}-\overrightarrow{x}^{\prime}) &  =\phi
_{m}(\overrightarrow{x}-(\overrightarrow{x}^{\prime}+\overrightarrow{d}%
))-\phi_{m}(\overrightarrow{x}-\overrightarrow{x}^{\prime})\nonumber\\
&  \simeq-\phi_{m}(\overrightarrow{x}-\overrightarrow{x}^{\prime
})\overrightarrow{d}.\frac{(\overrightarrow{x}-\overrightarrow{x}^{\prime}%
)}{|\overrightarrow{x}-\overrightarrow{x}^{\prime}|}(\frac{1}%
{|\overrightarrow{x}-\overrightarrow{x}^{\prime}|}+\sqrt{\frac{\sigma^{ISU}%
}{\epsilon\epsilon_{0}\text{ }D}}).
\end{align}
That is  a less symmetrical solution corresponding to the stationary response
of the medium to implant two 1 C point charges, one positive and another
negative, separated a distance $\overrightarrow{d}$ between them. The spatial
variation of the potential with the distance $\overrightarrow{y}$
$=\overrightarrow{x}-\overrightarrow{x}^{\prime}$ between the generation point
$\overrightarrow{x}^{\prime}$and the observation point $\overrightarrow{x}$ is
depicted in figure \ref{fig1} for two extremal values of the screening
distance $\lambda=$ $\sqrt{\frac{\epsilon\epsilon_{0}\text{ }D}{\sigma^{ISU}}%
}.$ The minimum value was chosen by calculating $\lambda$ for the
maximal value of the ISU conductivity $\sigma_{\max}^{ISU}$ in the range of
its typical values for tissues
\begin{equation}
\sigma_{\min}^{ISU}=0.05\text{ }\frac{\text{S}}{m}\leq\sigma^{ISU}\leq
\sigma_{\max}^{ISU}=0.3\text{ }\frac{\text{S}}{m},\label{1}%
\end{equation}
by also using the minimal values of the dielectric and the diffusion constants
within the ranges of their typical values
\begin{align}
\epsilon_{\min} &  =5\text{ }10^{5}\leq\epsilon\leq\epsilon_{\max}=8\text{
}10^{7},\label{2}\\
D_{\min} &  =0.45\times10^{-9}\leq D\leq D_{\max}=1.33\times10^{-9}\label{3}%
\end{align}

This defines the minimal value of $\lambda$ as%
\begin{align}
\lambda_{\min}  &  =\sqrt{\frac{\epsilon_{\min}\epsilon_{0}\text{ }D_{\min}%
}{\sigma_{\max}^{ISU}}}\nonumber\\
&  =8.14902\times10^{-8}\text{ \ }m.
\end{align}
In a similar way the maximal value of $\lambda$ becomes
\begin{align}
\lambda_{\max}  &  =\sqrt{\frac{\epsilon_{\max}\epsilon_{0}\text{ }D_{\max}%
}{\sigma_{\min}^{ISU}}}\nonumber\\
&  =4.34071\times10^{-6}\text{ }m.
\end{align}
The two extremal values of $\lambda$ illustrate that for the whole of set of
tissues showing the above range of data, the electrostatic fields have decay
lengths of the order of $4$ $\mu\, m$ at most. The figure \ref{fig1} shows
$\phi_{m}(\overrightarrow{x}-\overrightarrow{x}^{\prime})$ as function of the
observation distance $\ $measured in meters, for the larger value of
the decay distance $\lambda_{\max}=4.34071\times10^{-6}$ $\ m.$

For the comparison between $\phi_{m}$ and $\phi_{d}$ let us consider that the
vector $\overrightarrow{d}$ is parallel with the observation one
$(\overrightarrow{x}-\overrightarrow{x}^{\prime}).$ Thus the ratio of the two
potential signals become%
\begin{align}
|\frac{\phi_{d}(y)}{\phi_{m}(y)}|  &  =d\text{ }(\frac{1}{y}+\sqrt
{\frac{\sigma^{ISU}}{\epsilon\epsilon_{0}\text{ }D}}),\\
y  &  =|\overrightarrow{x}-\overrightarrow{x}^{\prime}|.
\end{align}

\begin{figure}[h]
\begin{center}
\includegraphics[width=15cm]{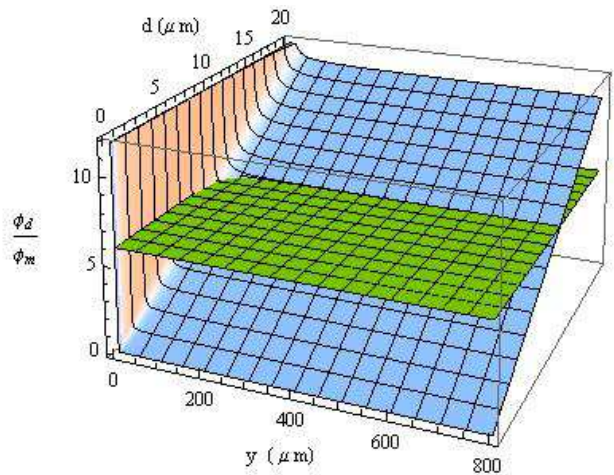}
\end{center}
\caption{ The plot shows the ratio the dipole and monopole potentials as a
function of the measuring distance and the sizes $d$ of the dipole. The
horizontal plane is plotted at the value equal to 6 of the estimated ratio
between the dipole and monopole measured potentials in reference
\cite{riera1}. It follows that mainly in all the region $d<10\,\mu\,m$ the
monopole to dipole ratio $\frac{\phi_{m}}{\phi_{m}}$ tends to increase with
respect to the value 1/6 estimated in \cite{riera1}. It can be observed that a
normal dendrite width is $1\,\mu\,m$. Thus, the impressed charge densities can
be seen as a superposition of densities of this typical width of $1\,\mu\,m$.
The evaluation again was done for a tissue showing a largest values of the
dielectric and diffusion constants and the lower conductivity value, in the
ranges defined in (\ref{1}),(\ref{2}),(\ref{3}). }%
\label{fig2}%
\end{figure}This ratio is plotted in figure \ref{fig2} as a function of the
size of the dipole $d$ and the distance $y$ between its position to the
measurement point. The parameters selected for this plot were
\begin{align}
D_{\max}  &  =1.33\times10^{-9},\\
\epsilon_{\max}  &  =8\text{ }10^{7},\\
\sigma_{\min}^{ISU}  &  =0.05.
\end{align}

They corresponds to the larger value of the length $\lambda_{\max}=\sqrt
{\frac{\epsilon_{\max}\epsilon_{0}\text{ }d_{\max}}{\sigma_{\min}^{ISU}}%
}=4.34071\times10^{-6}$ generated a material showing values of the
conductivity the dielectric and the diffusion constants restricted to be in
the usual experimental ranges (\ref{1}),(\ref{2}),(\ref{3}) for biological
tissues. The range of values for dipole size $d$ (which we interpret as
qualitatively representing the size of the region in which the impressed
currents are defined) in the plot of \ref{fig2}, runs from zero to $20$ $\mu
m$. The other axis is related with the distance at which the potential is
measured $y$. The range of values for $y$ of this axis is similar to the
spatial extension of the array of electrodes employed in \cite{riera1}.

\ The horizontal plane depicted in figure \ref{fig2} is plotted at a value of
$|\frac{\phi_{d}}{\phi_{m}}|=6$. This number approximately describes the ratio
between the amount of electric signals produced by pyramidal cells and the
corresponding amount defined by spin stellate cells, at the array of detectors
employed in reference \cite{riera1} (See figure 2 in that reference). The
rising with $d$ surface defines the values of $\ |\frac{\phi_{d}}{\phi_{m}}|$
as a function of $d,$ which we interpret as representing the size of non
symmetrical impressed sources; and as a function of the distance $y$ between
the point in which these sources are localized and the observation points at
the electrodes. These electrodes are separated by a 50 $\mu\,m$ in the
experiments done in \cite{riera1}. Then, the intersection curve between the
plane and plotted surface corresponds to pairs $(y,d),$ of measuring distances
$y$ and dimensions of the impressed sources $d$, at which the ratio between
the dipole and monopole signals shows the value observed in the experiments
done in \cite{riera1}. Thus, it follows that for sources sizes in the range
0-10 $\mu\,m$ the dipole potential has relative values with respect to the
monopole one, being lower than the number 6, which was estimated from the
results in \cite{riera1}. It can be noticed that the radius of the dendrites
taken to be of the order of 1 $\mu\,m$ makes reasonable to consider regions of
definition of impressed currents $d$ ranging from 1 to 10 $\mu\,m.$\

\ Therefore, the results in this work support the possibility for the
observation of monopole components of the impressed current when the
measurements are done in small regions of micrometer sizes.

\section{Summary}

We investigated the Maxwell equations for an homogeneous medium in which
electric currents are composed of Ohmic, diffusive and impressed currents. The
static approximation in which displacements currents are neglected was
employed. Solutions of the metallic charge density $\rho,$ electric field
potential and the magnetic field intensity as functions of the impressed
currents and densities ($\overrightarrow{J}_{imp}$, $\rho_{imp}$) are derived.
In addition, approximate formulae for these quantities are also obtained. The
analysis reproduce the conclusions of a previous work, by also indicating the
relevance of Yukawa like potentials. This representation indicates that for
spatial regions of reduced dimensions, the experiments can be able to detect
monopole like signals. The spatial sizes of the regions in which this turn to
be possible depend on a single parameter: the Yukawa decay length $\lambda=$
$\sqrt{\frac{\epsilon\epsilon_{0}\text{ }D}{\sigma^{ISU}}}$, which grows for
larger values of the relative dielectric constant and the diffusion parameter
and decreases for larger conductivities of the tissue medium. It also argued
that the magnetic tomographic signals result to be independent of the
existence of diffusion in the medium. The analysis also allowed to study the
ranges of spatial distances in which monopole signals can be detected.

\section*{Acknowledgment}

A.C.M. would like to acknowledge the support granted by the N-35 OEA Network
of the ICTP, Trieste, Italy.

\end{document}